\documentclass[onecolumn,preprint,prd,groupaddress,
showpacs,floatfix,nofootinbib]{revtex4}

\usepackage{epsfig}
\usepackage{amsmath,bm}

\begin{document}

\title{Relatively large $\theta_{13}$ and nearly maximal $\theta_{23}$ from the approximate
\\ $S_3$ symmetry of lepton mass matrices}

\author{Shun Zhou}

\email{zhoush@mppmu.mpg.de}

\affiliation{Max-Planck-Institut f\"ur Physik
(Werner-Heisenberg-Institut), F\"ohringer Ring 6, 80805 M\"unchen,
Germany}

\date{June 23, 2011}
\preprint{MPP-2011-74}

%%%%%%%%%%%%%%%%%%%%%%%%%%%%%%%%%%%%%%%%%%%%%%%%%%%%%%%%%%%%%%%%%%%%%%
\begin{abstract}
We apply the permutation symmetry $S_3$ to both charged-lepton and
neutrino mass matrices, and suggest a useful symmetry-breaking
scheme, in which the flavor symmetry is explicitly broken down via
$S_3 \to Z_3 \to \emptyset$ in the charged-lepton sector and via
$S_3 \to Z_2 \to \emptyset$ in the neutrino sector. Such a two-stage
breaking scenario is reasonable in the sense that both $Z_3$ and
$Z_2$ are the subgroups of $S_3$, while $Z_3$ and $Z_2$ only have a
trivial subgroup. In this scenario, we can obtain a relatively large
value of the smallest neutrino mixing angle, e.g., $\theta_{13}
\approx 9^\circ$, which is compatible with the recent result from
T2K experiment and will be precisely measured in the ongoing Double
Chooz and Daya Bay reactor neutrino experiments. Moreover, the
maximal atmospheric mixing angle $\theta_{23} \approx 45^\circ$ can
also be obtained while the best-fit value of solar mixing angle
$\theta_{12} \approx 34^\circ$ is assumed, which cannot be achieved
in previous $S_3$ symmetry models.
\end{abstract}
%%%%%%%%%%%%%%%%%%%%%%%%%%%%%%%%%%%%%%%%%%%%%%%%%%%%%%%%%%%%%%%%%%%%%%


\maketitle

\section{Introduction}

Flavor symmetry is currently a promising and widely-adopted approach
to understanding lepton mass spectra and neutrino mixing pattern
\cite{Ishimori:2010au}. In particular, a lot of attention has
recently been paid to discrete flavor symmetries, such as $A_4$
\cite{Ma:2001dn,Ma:2002ge,Altarelli:2005yp,Altarelli:2005yx} and
$S_4$ \cite{Yamanaka:1981pa, Brown:1984dk, Brown:1984mq,
Ma:2005pd,Hagedorn:2006ug,Zhang:2006fv,Lam:2008sh,Ding:2009iy,Yang:2011fh},
which are able to predict the tri-bimaximal neutrino mixing with
$\theta^{\rm T}_{12} = 35.3^\circ$, $\theta^{\rm T}_{23} = 45^\circ$
and $\theta^{\rm T}_{13} = 0$ that is well compatible with neutrino
oscillation experiments \cite{Harrison:2002er, Xing:2002sw,
Harrison:2002kp, He:2003rm}. However, the latest result from T2K
experiment indicates that $\theta_{13}$ is likely to be not
vanishing but relatively large. At the $90\%$ confidence level, the
T2K data are consistent with
\begin{equation}
5.0^\circ \lesssim \theta_{13} \lesssim 16.0^\circ \; ,
\end{equation}
in the case of normal neutrino mass hierarchy; and
\begin{equation}
5.8^\circ \lesssim \theta_{13} \lesssim 17.8^\circ \; ,
\end{equation}
in the case of inverted neutrino mass hierarchy, for a vanishing
Dirac CP-violating phase $\delta = 0$ \cite{Abe:2011sj}. As a matter
of fact, the global-fit analyses of neutrino oscillation experiments
before the T2K result have already shown some hint on a nonzero
$\theta_{13}$ \cite{Fogli:2008jx, GonzalezGarcia:2010er,
Schwetz:2011qt}. For instance, the latest best-fit values of three
neutrino mixing angles are $\theta_{12} = 34^\circ$, $\theta_{23} =
46^\circ$ and $\theta_{13} = 6^\circ$ \cite{Schwetz:2011qt}. No
doubt the symmetry-breaking terms in the $A_4$ or $S_4$ models can
account for a relatively large $\theta_{13}$, but one has to avoid
the resultant large corrections to $\theta^{\rm T}_{12}$ and
$\theta^{\rm T}_{23}$, which are already in excellent agreement with
experimental data
\cite{Shimizu:2011xg,King:2011xu,Ma:2011yi,He:2011gb}.

Therefore, we are well motivated to consider the simplest
non-Abelian discrete symmetry $S_3$ for lepton mass matrices
\cite{Fritzsch:1995dj, Fukugita:1998vn, Koide:1999mx,
Fritzsch:1998xs, Fritzsch:1999im, Tanimoto:2000fz, Branco:2001hn,
Fujii:2002jw, Kubo:2003iw, Chen:2004rr, Fritzsch:2004xc,
Grimus:2005mu, Mohapatra:2006pu, Jora:2006dh, Chen:2007zj,
Jora:2009gz, Dicus:2010iq, Xing:2010iu, Jora:2010at,Dev:2011qy}. A
salient feature of the $S_3$ model is the prediction of democratic
neutrino mixing pattern \cite{Fritzsch:1995dj, Fritzsch:1998xs} with
$\theta^{\rm D}_{12} = 45^\circ$, $\theta^{\rm D}_{23} = 54.7^\circ$
and $\theta^{\rm D}_{13} = 0$, which is now disfavored by current
neutrino oscillation data. As argued in Refs. \cite{Xing:2010pn,
Xing:2011at}, however, significant corrections from the
symmetry-breaking terms may modify $\theta^{\rm D}_{12}$ and
$\theta^{\rm D}_{23}$ to be consistent with the observed values, and
simultaneously give rise to a relatively large $\theta_{13}$. This
observation is indeed intriguing because the perturbations to the
symmetry-limit values of three mixing angles are comparable in
magnitude.

In this paper, we reconsider the $S_3$ symmetry and its explicit
breaking for lepton mass matrices, and demonstrate that a relatively
large $\theta_{13}$ can be achieved while both $\theta_{12}$ and
$\theta_{23}$ are in good agreement with neutrino oscillation
experiments. Note that we shall follow a phenomenological approach
and work at the mass-matrix level, however, the derived patterns of
lepton mass matrices and the proposed symmetry-breaking scheme may
be helpful for the model building at the field-theory level. Our
work differs from previous ones in several aspects. First, we apply
the $S_3$ symmetry to both charged-lepton and neutrino mass
matrices. In Refs. \cite{Chen:2007zj, Jora:2009gz, Jora:2010at}, the
$S_3$ symmetry has only been applied to the neutrino mass matrix,
and the exactly or nearly tri-bimaximal neutrino mixing can be
derived. Second, we propose an interesting symmetry-breaking scheme,
i.e., $S_3 \to Z_3 \to \emptyset$ for charged leptons and $S_3 \to
Z_2 \to \emptyset$ for neutrinos. Such a two-stage breaking scenario
is quite natural, because both $Z_3$ and $Z_2$ are the subgroups of
$S_3$, while $Z_3$ and $Z_2$ only have a trivial subgroup. As a
consequence of this breaking scheme, the charged-lepton mass matrix
is non-symmetric, while neutrino mass matrix is still symmetric as
it should be, because neutrinos are assumed to be Majorana
particles. Third, we can get both a relatively large $\theta_{13}$
and a nearly maximal $\theta_{23}$, which cannot be reached in the
previous $S_3$ models \cite{Fritzsch:2004xc,Xing:2010iu}.

In Sec. II, the lepton mass matrices in the $S_3$-symmetry limit and
the symmetry-breaking terms are constructed. The phenomenological
implications for lepton mass spectra and neutrino mixing angles are
explored in Sec. III. We summarize our conclusions in Sec. IV.

\section{$S_3$ Symmetry and Its Breaking}

From the phenomenological point of view, the lepton masses and
mixing angles at low energies are determined by lepton mass terms
\begin{equation}
-{\cal L}_{\rm m} = \overline{\ell_{\rm L}} M_{\ell} \ell_{\rm R} +
\frac{1}{2} \overline{\nu_{\rm L}} M_\nu \nu^c_{\rm L} + {\rm h.c.}
\; ,
\end{equation}
where $M_\ell$ stands for the mass matrix of charged leptons, and
$M_\nu$ for the effective mass matrix of Majorana neutrinos. The
latter can be realized in various neutrino mass models, such as
seesaw models, which extend the standard model by introducing
singlet or triplet fermions, or triplet scalars \cite{Xing:2011zza}.

As usual, we can decompose the lepton mass matrices into a
symmetry-limit part and a symmetry-breaking perturbation term:
\begin{equation}
M_l = M^{(0)}_l + \Delta M_l \; , ~~~~~ M_\nu = M^{(0)}_\nu + \Delta
M_\nu \; .
\end{equation}
In the $S_3$-symmetry limit, the Lagrangian in Eq. (3) is invariant
under the transformation $\ell_{\rm L} \to S^{(ijk)} \ell_{\rm L}$,
$\ell_{\rm R} \to S^{(ijk)} \ell_{\rm R}$ and $\nu_{\rm L} \to
S^{(ijk)} \nu_{\rm L}$ with $S^{(ijk)}$ being the group elements of
$S_3$. The three-dimensional representations of all six group
elements are
\begin{eqnarray}
&& S^{(123)} = \left(\begin{matrix}1 & 0 & 0 \cr 0 & 1 & 0 \cr 0 & 0
& 1 \end{matrix}\right) \; , ~~~~ S^{(231)} = \left(\begin{matrix}0
& 1 & 0 \cr 0 & 0 & 1 \cr 1 & 0 & 0 \end{matrix}\right) \; ,
\nonumber \\
&& S^{(312)} = \left(\begin{matrix}0 & 0 & 1 \cr 1 & 0 & 0 \cr 0 & 1
& 0\end{matrix}\right) \; , ~~~~ S^{(213)} = \left(\begin{matrix}0 &
1 & 0 \cr 1 & 0 & 0 \cr 0 & 0 & 1
\end{matrix}\right) \; , \nonumber \\
&& S^{(132)} = \left(\begin{matrix}1 & 0 & 0 \cr 0 & 0 & 1 \cr 0 & 1
& 0 \end{matrix}\right) \; , ~~~~ S^{(321)} = \left(\begin{matrix}0
& 0 & 1 \cr 0 & 1 & 0 \cr 1 & 0 & 0\end{matrix}\right) \; .
\end{eqnarray}
Thus $M^{(0)}_\ell$ and $M^{(0)}_\nu$ should commutate with
$S^{(ijk)}$, i.e., $[M^{(0)}_\ell, S^{(ijk)}] =0$ and $[M^{(0)}_\nu,
S^{(ijk)}] = 0$. The most general form of $M^{(0)}_\ell$ and
$M^{(0)}_\nu$ with $S_3$ symmetry is \cite{Xing:2010iu}
\begin{eqnarray}
M^{(0)}_\ell &=& \frac{c_\ell}{3} \left[\left(\begin{matrix} 1 & 1 &
1 \cr 1 & 1 & 1 \cr  1 & 1 & 1 \end{matrix}\right) + r_\ell
\left(\begin{matrix} 1 & 0 & 0 \cr 0 & 1 & 0 \cr  0 & 0 & 1
\end{matrix}\right) \right] \; , \nonumber \\
M^{(0)}_\nu &=& c_\nu \left[\left(\begin{matrix} 1 & 0 & 0 \cr 0 & 1
& 0 \cr  0 & 0 & 1 \end{matrix}\right) + r_\nu \left(\begin{matrix}
1 & 1 & 1 \cr 1 & 1 & 1 \cr  1 & 1 & 1
\end{matrix}\right) \right] \; ,
\end{eqnarray}
where the real and positive parameters $c_\ell$ and $c_\nu$ set the
mass scales of charged leptons and neutrinos, respectively. Since
both $M^{(0)}_\ell$ and $M^{(0)}_\nu$ can be diagonalized by the
same orthogonal matrix
\begin{equation}
V_{\rm D} = \frac{1}{\sqrt{6}} \left(\begin{matrix} \sqrt{3} & 1 &
\sqrt{2} \cr -\sqrt{3} & 1 & \sqrt{2} \cr 0 & -2 & \sqrt{2}
\end{matrix}\right) \; ,
\end{equation}
the neutrino mixing matrix turns out to be an identity matrix. In
other words, the democratic mixing arising from the charged-lepton
sector gets too large corrections from the neutrino sector, or vice
versa. Additionally, the first two generations of leptons are
exactly degenerate in mass. In the limit of $r_\nu = 0$, the
neutrino mass matrix is diagonal and we obtain the democratic
mixing. But neutrinos are exactly degenerate in mass. The
perturbation terms explicitly breaking the $S_3$ symmetry are
necessary to generate realistic lepton mass spectra and neutrino
mixing angles.

Note that the group elements of $S_3$ can be categorized into three
conjugacy classes ${\cal C}_0 = \left\{S^{(123)}\right\}$, ${\cal
C}_1 = \left\{S^{(231)}, S^{(312)}\right\}$ and ${\cal C}_2 =
\left\{S^{(213)}, S^{(132)}, S^{(321)}\right\}$. It is
straightforward to show that the invariant subgroup of $S_3$ is the
cyclic group of order three
\begin{equation}
Z_3 = \left\{S^{(123)}, S^{(231)}, S^{(312)}\right\} \equiv
\left\{e, a, a^2\right\} \; ,
\end{equation}
where we have defined the identity element as $e \equiv S^{(123)}$
and the generator of $Z_3$ group as $a \equiv S^{(231)}$. With the
explicit representations in Eq. (5), one can immediately verify $a^3
= e$. The $S_3$ group has three $Z_2$ subgroups
\begin{eqnarray}
Z^{(12)}_2 &=& \left\{S^{(123)}, S^{(213)}\right\} \; , \nonumber \\
Z^{(23)}_2 &=& \left\{S^{(123)}, S^{(132)}\right\} \; , \nonumber \\
Z^{(31)}_2 &=& \left\{S^{(123)}, S^{(321)}\right\} \; .
\end{eqnarray}
The $Z^{(23)}_2$ group can be identified with the $\mu$-$\tau$
symmetry, which has been extensively discussed in connection with
the maximal atmospheric mixing angle and the small reactor mixing
angle \cite{Fukuyama:1997ky,Xing:2006xa}.

Obviously, it is natural to explicitly break $S_3$ symmetry to its
subgroups. Along this line, we propose to construct the perturbation
terms $\Delta M_\ell$ and $\Delta M_\nu$ as
\begin{eqnarray}
\Delta M_\ell &=& \Delta M^{(1)}_\ell + \Delta M^{(2)}_\ell \; ,
\nonumber \\
\Delta M_\nu &=&  \Delta M^{(1)}_\nu + \Delta M^{(2)}_\nu \; ,
\end{eqnarray}
such that the flavor symmetry is explicitly broken down via the
chain
\begin{equation}
\begin{array}{ccccc}
  ~ & \Delta M^{(1)}_\ell & ~ & \Delta M^{(2)}_\ell & ~ \\
  S_3 & \longrightarrow & Z_3 & \longrightarrow & \emptyset
\end{array}
\end{equation}
in the charged-lepton sector, while via a distinct chain
\begin{equation}
\begin{array}{ccccc}
  ~ & \Delta M^{(1)}_\nu & ~ & \Delta M^{(2)}_\nu & ~ \\
  S_3 & \longrightarrow & Z_2 & \longrightarrow & \emptyset
\end{array}
\end{equation}
in the neutrino sector. It is worthwhile to remark that the mass
term breaking $S_3$ to $Z_3$ is proportional to $S^{(231)}$ or
$S^{(312)}$, which is a non-symmetric matrix, thus the
symmetry-breaking chain in Eq. (11) is only allowed for charged
leptons. Neutrino mass matrix must be symmetric because we have
assumed neutrinos to be Majorana particles as in a class of seesaw
models.

It is easy to show that $\Delta M^{(1)}_\ell$ can always be cast
into the following form
\begin{equation}
\Delta M^{(1)}_\ell = \frac{c_\ell}{3} \left(\begin{matrix} 0 &
\delta_\ell & 0 \cr 0 & 0 & \delta_\ell \cr  \delta_\ell & 0 & 0
\end{matrix}\right)
\end{equation}
by redefining the parameters $c_\ell$ and $r_\ell$ in
$M^{(0)}_\ell$. The first-order perturbation term in the neutrino
mass matrix can be written as
\begin{equation}
\Delta M^{(1)}_\nu = c_\nu \left(\begin{matrix} \delta_\nu & 0 & 0
\cr 0 & 0 & \delta_\nu \cr 0 & \delta_\nu & 0
\end{matrix}\right) \; ,
\end{equation}
which reduces the $S_3$ symmetry to $Z^{(23)}_2$ or $\mu$-$\tau$
symmetry. Note that here we take $Z^{(23)}_2$ for example, and one
can discuss similarly the other two possibilities $Z^{(12)}_2$ and
$Z^{(31)}_2$. Nevertheless, the second-stage perturbation terms
$\Delta M^{(2)}_\ell$ and $\Delta M^{(2)}_\nu$, which respectively
break down the residual $Z_3$ and $Z_2$ symmetries, could have many
different forms. As both of them are intended for breaking the mass
degeneracy between the first and second generations, we choose the
diagonal form for simplicity \cite{Fritzsch:2004xc}
\begin{eqnarray}
\Delta M^{(2)}_\ell &=& \frac{c_\ell}{3} \left(\begin{matrix}
-i\epsilon_\ell & 0 & 0 \cr 0 & +i\epsilon_\ell & 0 \cr 0 & 0 &
+\varepsilon_\ell \end{matrix}\right) \; , \nonumber \\
\Delta M^{(2)}_\nu &=& c_\nu \left(\begin{matrix} -\epsilon_\nu & 0
& 0 \cr 0 & +\epsilon_\nu & 0  \cr 0 & 0 & +\varepsilon_\nu
\end{matrix}\right) \; .
\end{eqnarray}
Now that $\Delta M^{(2)}_\ell$ is complex, we expect the CP
violation in the lepton sector. Furthermore, all the parameters
$r_f$, $\delta_f$, $\epsilon_f$ and $\varepsilon_f$ for $f = \ell,
\nu$ are assumed to be real and serve as small perturbations, i.e.,
$|r_f|, |\delta_f|, |\epsilon_f| \ll |\varepsilon_f| < 1$. At this
moment, we have completed the construction of the lepton mass
matrices in Eq. (4).

\section{Lepton Masses and Mixing Angles}

Now we are ready to figure out the lepton mass spectra and neutrino
mixing angles. In general, the charged-lepton mass matrix $M_\ell$
is an arbitrary complex matrix and can be diagonalized by a
bi-unitary transformation $U^\dagger_\ell M_\ell \tilde{U}_\ell =
{\rm Diag}\{m_e, m_\mu, m_\tau\}$, where $m_\alpha$ (for $\alpha =
e, \mu, \tau$) are charged-lepton masses and the matrices $U_\ell$
and $\tilde{U}_\ell$ are unitary. Since $\tilde{U}_ \ell$ is
associated with the right-handed fields of charged leptons and has
nothing to do with the lepton flavor mixing, it is more convenient
to consider the Hermitian matrix $H_\ell \equiv M_\ell
M^\dagger_\ell$, which can be diagonalized as $U^\dagger_\ell H_\ell
U_\ell = {\rm Diag}\{m^2_e, m^2_\mu, m^2_\tau\}$. Furthermore, we
shall work in the so-called hierarchy basis, where the relevant
matrix is $H^\prime_\ell \equiv V^T_{\rm D} H_\ell V_{\rm D}$. In
the leading-order approximation, we arrive at
\begin{widetext}
\begin{equation}
H^\prime_\ell = \frac{c^2_\ell}{9} \left(\begin{matrix} r^2_\ell
-r_\ell \delta_\ell + \delta^2_\ell + \epsilon^2_\ell  &
\displaystyle \frac{\delta_\ell \varepsilon_\ell}{\sqrt{3}} &
-i\sqrt{6} \epsilon_\ell \cr \displaystyle \frac{\delta_\ell
\varepsilon_\ell}{\sqrt{3}} & \displaystyle
\frac{2}{3}\varepsilon_\ell (\varepsilon_\ell + 2 r_\ell -
\delta_\ell) & -\sqrt{2} \varepsilon_\ell \cr i\sqrt{6}
\epsilon_\ell & -\sqrt{2} \varepsilon_\ell & 9 + 2\varepsilon_\ell
\end{matrix}\right)
\end{equation}
\end{widetext}
with a rational assumption of $|r_\ell|, |\delta_\ell|,
|\epsilon_\ell| \ll |\varepsilon_\ell| < 1$. After diagonalizing the
above matrix via $V^\dagger_\ell H^\prime V_\ell = {\rm
Diag}\{m^2_e, m^2_\mu, m^2_\tau\}$, one obtains three charged-lepton
masses
\begin{eqnarray}
m_e &\approx& c_\ell \left|\frac{r_\ell}{3} - \frac{\delta_\ell}{6}
+ \frac{\epsilon^2_\ell}{6\varepsilon_\ell} +
\frac{3\delta^2_\ell}{8\varepsilon_\ell}\right|
\; , \nonumber \\
m_\mu &\approx& c_\ell \left(\frac{2}{9} \varepsilon_\ell +
\frac{1}{3}
r_\ell - \frac{1}{6}\delta_\ell \right) \; , \\
m_\tau &\approx& c_\ell \left(1 + \frac{1}{9} \varepsilon_\ell +
\frac{1}{3} r_\ell + \frac{1}{3} \delta_\ell\right)\; . \nonumber
\end{eqnarray}
Defining $m_0 = c_\ell (2r_\ell - \delta_\ell)/6$, we have $|m_0| <
m_\mu$. The small parameters $\varepsilon_\ell$, $\delta_\ell$ and
$\epsilon_\ell$ can be expressed in terms of charged-lepton masses
and the $m_0$ parameter
\begin{equation}
\varepsilon_\ell \approx \frac{9}{2} \frac{m_\mu - m_0}{m_\tau -
m_0} \; , ~~~~ \frac{\epsilon^2_\ell}{\varepsilon^2_\ell} +
\frac{9\delta^2_\ell}{4\varepsilon^2_\ell} \approx \frac{4}{3}
\frac{|m_e - |m_0||}{m_\mu - m_0} \; .
\end{equation}
For $\delta_\ell = 0$, one immediately reproduces the same results
in Ref. \cite{Xing:2010iu}, where the perturbation term $\Delta
M^{(1)}_\ell$ is absent in the charged-lepton mass matrix. The
unitary matrix $U_\ell = V_{\rm D} V_\ell$ is found to be
\begin{widetext}
\begin{eqnarray}
U_\ell \approx V_{\rm D} + \frac{e^{-i\phi}}{\sqrt{6}}
\frac{\sqrt{|m_e - |m_0||}}{\sqrt{m_\mu - m_0}} \left(\begin{matrix}
1 & \sqrt{3} & 0 \cr 1 & -\sqrt{3} & 0 \cr -2 & 0 & 0
\end{matrix}\right) + \frac{1}{2\sqrt{3}}
\frac{m_\mu - m_0}{m_\tau - m_0} \left(\begin{matrix} 0 & \sqrt{2} &
-1 \cr 0 & \sqrt{2} & -1 \cr 0 & \sqrt{2} & 2
\end{matrix}\right) \; ,
\end{eqnarray}
\end{widetext}
where $\phi \equiv \arctan[2\epsilon_\ell/(3\delta_\ell)]$ gives
rise to the Dirac CP-violating phase. Comparing Eq. (19) with the
counterpart in Ref. \cite{Xing:2010iu}, we can observe that the
additional symmetry-breaking term $\Delta M^{(1)}_\ell$ or the
parameter $\delta_\ell$ can be determined by measuring the CP
violation in neutrino oscillations, which is indeed to be performed
in the long-baseline neutrino experiments. It is worth mentioning
that there are five real parameters in the charged-lepton mass
matrix (i.e., $c_\ell$, $r_\ell$, $\delta_\ell$, $\epsilon_\ell$ and
$\varepsilon_\ell$), which can be expressed in terms of
charged-lepton masses $(m_e, m_\mu, m_\tau)$ and $(m_0, \phi)$. The
latter two enter into the neutrino mixing matrix, and can be
determined by neutrino mixing angles and the Dirac CP-violating
phase, as we shall show later.

Next, we turn to the neutrino mass matrix given in Eqs. (6), (14)
and (15). The unitary matrix $U_\nu$ used to diagonalize $M_\nu$
through $U^\dagger_\nu M_\nu U^*_\nu = {\rm Diag}\{m_1, m_2, m_3\}$
is approximately given by \cite{Fritzsch:2004xc,Xing:2010iu}
\begin{equation}
U_\nu \approx \frac{1}{\varepsilon_\nu} \left( \begin{matrix}
\varepsilon_\nu c_\theta & \varepsilon_\nu s_\theta & r_\nu \cr
-\varepsilon_\nu s_\theta & \varepsilon_\nu c_\theta & r_\nu +
\delta_\nu \cr (r_\nu + \delta_\nu)s_\theta - r_\nu c_\theta &
-(r_\nu + \delta_\nu) c_\theta - r_\nu s_\theta & \varepsilon_\nu
\end{matrix} \right) \; ,
\end{equation}
where $c_\theta \equiv \cos \theta$ and $s_\theta = \sin \theta$
with $\tan 2\theta \equiv 2r_\nu/(2\epsilon_\nu - \delta_\nu)$. Note
that the perturbation parameters satisfy $r_\nu, \delta_\nu,
\epsilon_\nu \ll \varepsilon_\nu < 1$ as in the case of charged
leptons. Three neutrino mass eigenvalues are
\begin{eqnarray}
m_3 &\approx& c_\nu \left(1 + r_\nu + \varepsilon_\nu\right) \; ,
\nonumber \\
m_2 &\approx& c_\nu \left(1 + r_\nu + \delta_\nu/2 +
\sqrt{(\epsilon_\nu - \delta_\nu/2)^2 + r^2_\nu}\right) \; ,
\\
m_1 &\approx& c_\nu \left(1 + r_\nu + \delta_\nu/2 -
\sqrt{(\epsilon_\nu - \delta_\nu/2)^2 + r^2_\nu}\right) \; ,
\nonumber
\end{eqnarray}
where we have assumed the normal mass hierarchy. It is
straightforward to calculate the neutrino mass-square differences
$\Delta m^2_{31} \approx 2c^2_\nu \varepsilon_\nu$ and $\Delta
m^2_{21} \approx 2c^2_\nu \sqrt{(2\epsilon_\nu - \delta_\nu)^2 +
4r^2_\nu}$, for which the latest best-fit values are $\Delta
m^2_{21} = 7.59\times 10^{-5}~{\rm eV}^2$ and $\Delta m^2_{31} =
2.45\times 10^{-3}~{\rm eV}^2$ \cite{Schwetz:2011qt}. As indicated
by Eq. (21), we have nearly degenerate neutrino masses. Therefore,
the effective neutrino mass in tritium beta decays $\langle m_\beta
\rangle$ and that in neutrinoless double-beta decays $\langle
m_{\beta \beta} \rangle$ are on the same order of the mass-scale
parameter $c_\nu$. Currently, the most stringent bound $\langle
m_\beta \rangle \approx \langle m_{\beta \beta} \rangle \approx
c_\nu \sim {\cal O}(0.1~{\rm eV})$ comes from cosmological
observations \cite{pdg}. With the help of neutrino mass-squared
differences, we can estimate \cite{Fritzsch:2004xc}
\begin{equation}
\varepsilon_\nu \approx \frac{\Delta m^2_{31}}{2\langle m_\beta
\rangle^2} \approx 0.12\; , ~~~ \sqrt{(2\epsilon_\nu - \delta_\nu)^2
+ 4r^2_\nu} \approx \frac{\Delta m^2_{21}}{2\langle m_\beta
\rangle^2} \approx 3.8 \times 10^{-3} \; .
\end{equation}
In order to further fix the model parameters $r_\nu$, $\delta_\nu$
and $\epsilon_\nu$, we have to study neutrino mixing angles and the
Dirac CP-violating phase.

From Eqs. (19) and (20), we can derive the neutrino mixing matrix,
which is defined as $V \equiv U^\dagger_\ell U_\nu$. More
explicitly,
\begin{widetext}
\begin{eqnarray}
V &\approx& \frac{1}{\sqrt{6}} \left(\begin{matrix}
\sqrt{3}(c_\theta + s_\theta) & -\sqrt{3}(c_\theta - s_\theta) & 0
\cr (c_\theta - s_\theta) & (c_\theta + s_\theta) & -2 \cr
\sqrt{2}(c_\theta - s_\theta) & \sqrt{2}(c_\theta + s_\theta) &
\sqrt{2} \end{matrix}\right) \nonumber \\
&& + \frac{1}{2\sqrt{3}} \frac{m_\mu - m_0}{m_\tau - m_0}
\left(\begin{matrix} 0 & 0 & 0 \cr \sqrt{2}(c_\theta - s_\theta) &
\sqrt{2}(c_\theta + s_\theta) & \sqrt{2} \cr -(c_\theta - s_\theta)
& -(c_\theta + s_\theta) & 2
\end{matrix}\right) \nonumber \\
&& + \frac{e^{i\phi}}{\sqrt{6}} \frac{\sqrt{|m_e -
|m_0||}}{\sqrt{m_\mu - m_0}} \left(\begin{matrix} (c_\theta -
s_\theta) & (c_\theta + s_\theta) & -2 \cr \sqrt{3}(c_\theta +
s_\theta) & -\sqrt{3}(c_\theta - s_\theta) & 0 \cr 0 & 0 & 0
\end{matrix}\right) \nonumber \\
& & + \frac{1}{\sqrt{6}} \frac{r_\nu}{\varepsilon_\nu}
\left(\begin{matrix} 0 & 0 & 0 \cr 2(c_\theta - s_\theta) &
2(c_\theta + s_\theta) & 2 \cr -\sqrt{2}(c_\theta - s_\theta) &
-\sqrt{2}(c_\theta + s_\theta) & 2\sqrt{2}
\end{matrix}\right)  + \frac{1}{\sqrt{6}} \frac{\delta_\nu}{\varepsilon_\nu}
\left(\begin{matrix} 0 & 0 & -\sqrt{3} \cr - 2s_\theta & 2c_\theta &
1 \cr \sqrt{2} s_\theta & -\sqrt{2} c_\theta & \sqrt{2}
\end{matrix}\right) \; ,~~~~~
\end{eqnarray}
\end{widetext}
where the last term arises from the symmetry-breaking term $\Delta
M^{(1)}_\nu$, which evidently contributes to both $\theta_{13}$ and
$\theta_{23}$. Comparing between Eq. (23) and the standard
parametrization of neutrino mixing matrix \cite{pdg}, one can
extract three neutrino mixing angles and the CP-violating phase.
Some comments are in order:
\begin{enumerate}
\item The solar mixing angle $\theta_{12}$ is determined by $
\sin^2 2\theta_{12} \approx 4|V_{e1}|^2 |V_{e2}|^2 \approx \cos^2
2\theta$ with $\tan 2\theta = 2r_\nu/(2\epsilon_\nu - \delta_\nu)$,
so the perturbation parameters $r_\nu$, $\epsilon_\nu$ and
$\delta_\nu$ should satisfy
\begin{equation}
\frac{2r_\nu}{\left|2\epsilon_\nu - \delta_\nu\right|} = \cot
2\theta_{12} = 0.4 \; ,
\end{equation}
where the best-fit value $\theta_{12} = 34^\circ$ has been input
\cite{Schwetz:2011qt}. Combining Eqs. (22) and (24), one can get
$r_\nu \approx 7.0\times 10^{-4} \ll \varepsilon_\nu$, which
justifies our assumption $r_\nu, \delta_\nu, \epsilon_\nu \ll
\varepsilon_\nu < 1$ for perturbation parameters. The ratio
$\epsilon_\nu/\delta_\nu$ is thus the only unfixed parameter in the
neutrino sector.

\item The smallest neutrino mixing angle $\theta_{13}$ is given by
\begin{equation}
\sin \theta_{13} \approx \left|\frac{2}{\sqrt{6}} e^{i\phi}
\frac{\sqrt{|m_e - |m_0||}}{\sqrt{m_\mu - m_0}} + \frac{1}{\sqrt{2}}
\frac{\delta_\nu}{\varepsilon_\nu}\right| \; ,
\end{equation}
while the Dirac CP-violating phase by
\begin{equation}
\delta \approx \arg\left[\frac{2}{\sqrt{6}} e^{i\phi}
\frac{\sqrt{|m_e - |m_0||}}{\sqrt{m_\mu - m_0}} + \frac{1}{\sqrt{2}}
\frac{\delta_\nu}{\varepsilon_\nu}\right] \; .
\end{equation}
Note that $\theta_{13}$ receives contributions both from
charged-lepton and neutrino sectors. If $m_0 = 0$ and $\delta_\nu =
0$ are taken, as in Ref. \cite{Fritzsch:2004xc}, we get $\sin
\theta_{13} \approx \sqrt{2m_e/3m_\mu} \approx 0.057$ or
$\theta_{13} \approx 3.2^\circ$ by inputting $m_e = 0.4866 ~{\rm
MeV}$ and $m_\mu = 102.718~{\rm MeV}$ at the electroweak scale
\cite{Xing:2007fb}. As observed in Ref. \cite{Xing:2010iu}, when
$m_0$ is switched on and set to $m_0 < 0$ and $|m_0| > m_e$, one can
get relatively large values of $\theta_{13}$ and saturate the upper
bound for $m_0 \approx - 14~m_e$. In our scenario, the sizable
$\theta_{13}$ can be obtained even for somewhat smaller $|m_0|$ due
to the $\delta_\nu/\varepsilon_\nu$ term.

\item The atmospheric mixing angle $\theta_{23}$ is given by
\begin{equation}
\sin 2\theta_{23} = \frac{2\sqrt{2}}{3} \left(1 + \frac{1}{2}
\frac{m_\mu - m_0}{m_\tau - m_0} + \frac{r_\nu}{\varepsilon_\nu} +
\frac{1}{2} \frac{\delta_\nu}{\varepsilon_\nu}\right) \; ,
\end{equation}
which can also be enhanced due to the symmetry-breaking term $\Delta
M^{(1)}_\nu$ or the $\delta_\nu$ parameter. If $\delta_\nu$ is
vanishing, one obtains $\sin 2\theta_{23} \approx 0.97$ or
$\theta_{23} \approx 38^\circ$ by inputting $m_\mu = 102.718~{\rm
MeV}$ and $m_\tau = 1746.24 ~{\rm MeV}$ at the electroweak scale
\cite{Xing:2007fb}. The nearly maximal mixing angle $\theta_{23}
\approx 45^\circ$ cannot be achieved even for $m_0 = -14~m_e$, which
is necessary to generate relatively-large $\theta_{13}$
\cite{Xing:2010iu}. As indicated by Eq. (27), however, $\theta_{23}$
can be nearly maximal in our scenario with a nonvanishing
$\delta_\nu$.
\end{enumerate}

To illustrate how our model can accommodate both relatively-large
$\theta_{13}$ and nearly-maximal $\theta_{23}$, we introduce $\xi
\equiv \epsilon_\nu/\delta_\nu$ and $\zeta \equiv |m_0|/m_e$, and
rewrite Eqs. (25) and (27) in terms of $(\phi, \xi, \zeta)$ and
physical observables
\begin{equation}
\sin \theta_{13} \approx \left|e^{i\phi} \sqrt{\frac{2|\zeta -
1|}{3(\zeta + m_\mu/m_e)}} + \frac{\Delta m^2_{21}}{\Delta m^2_{31}}
\frac{\sin 2\theta_{12}}{\sqrt{2}|2\xi - 1|}\right| \; ,
\end{equation}
and
\begin{equation}
\sin 2\theta_{23} \approx \frac{2\sqrt{2}}{3} \left[1 + \frac{1}{2}
\frac{m_\mu/m_e + \zeta}{m_\tau/m_e + \zeta} + \frac{\Delta
m^2_{21}}{\Delta m^2_{31}} \left(\frac{\sin 2\theta_{12}}{2|2\xi -
1|} + \frac{\cos 2\theta_{12}}{2}\right)\right] \; ,
\end{equation}
where $m_0 < 0$ and $\xi \ne 1/2$ have been assumed. Hence three
remaining parameters $(\phi, \xi, \zeta)$ are actually fixed by
$\delta$, $\theta_{13}$ and $\theta_{23}$, which can be measured in
neutrino oscillation experiments. Note that $|m_0| = |c_\ell(2r_\ell
-\delta_\ell)|/6$ is naturally on the same order of $m_e$, thus
$\sin 2\theta_{23}$ is insensitive to $\zeta$ because of the strong
mass hierarchy of charged leptons $m_\tau \gg m_\mu \gg m_e$. In
this case, we can safely neglect $\zeta$ in Eq. (29) and solve it
analytically for the $\xi$ parameter
\begin{equation}
\xi \approx \frac{1}{2} \pm \frac{\sin 2\theta_{12}}{\displaystyle
4\left[\left(\frac{3}{2\sqrt{2}} \sin 2\theta_{23} - \frac{1}{2}
\frac{m_\mu}{m_\tau} - 1\right)\frac{\Delta m^2_{31}}{\Delta
m^2_{21}} - \frac{\cos 2\theta_{12}}{2}\right]} \; ,
\end{equation}
where the upper and lower sign stands for $\xi > 1/2$ and $\xi <
1/2$, respectively. In assumption of $\phi = 0$, we can further
solve Eq. (28) for the $\zeta$ parameter
\begin{equation}
\zeta \approx 1 + \frac{3m_\mu}{2m_e} \left(\sin \theta_{13} -
\frac{3}{2} \sin 2\theta_{23} + \sqrt{2} + \frac{1}{\sqrt{2}}
\frac{m_\mu}{m_\tau} - \frac{\Delta m^2_{21}}{\Delta m^2_{31}}
\frac{\cos 2\theta_{12}}{\sqrt{2}}\right)^2 \; .
\end{equation}
In order to obtain $\theta_{13} \approx 9^\circ$ and $\theta_{23}
\approx 45^\circ$ as well, one can insert the best-fit values
$\Delta m^2_{21} = 7.59\times 10^{-5}~{\rm eV}^2$, $\Delta m^2_{31}
= 2.45\times 10^{-3}~{\rm eV}^2$ and $\theta_{12} = 34^\circ$
\cite{Schwetz:2011qt}, together with the charged-lepton masses, into
Eqs. (30) and (31), and finally find $\zeta \approx 5$ and $\xi
\approx 0.2$ or $\xi \approx 0.8$. In the neutrino sector, we can
estimate the model parameters as $r_\nu \approx 7.0\times 10^{-4}$,
$\delta_\nu \approx 5.8\times 10^{-3}$, $\epsilon_\nu \approx
1.16\times 10^{-3}$ for $\xi = 0.2$ or $\epsilon_\nu \approx
4.64\times 10^{-3}$ for $\xi = 0.8$, and $\varepsilon_\nu = 0.12$,
which are consistent with the requirement that $r_\nu, \delta_\nu,
\epsilon_\nu \ll \varepsilon_\nu$. In the charged-lepton sector, we
get $m_0 \approx - 5m_e$ from $\zeta \approx 5$, and have assumed
$\phi = 0$. The latter condition implies $\epsilon_\ell = 0$, and
thus one can find from Eq. (18) that $\varepsilon_\ell \approx
9m_\mu/(2m_\tau) \approx 0.26$, $\delta_\ell \approx 8\sqrt{m_e}
\varepsilon_\ell/(3\sqrt{3m_\mu}) \approx 0.03$ and $r_\ell \approx
\delta_\ell$, which are also in agreement with $r_\ell, \delta_\ell,
\epsilon_\ell \ll \varepsilon_\ell$. Therefore, both $\theta_{13}
\approx 9^\circ$ and $\theta_{23} \approx 45^\circ$ can indeed be
achieved in our scenario. For the case with $\phi \neq 0$, we can
completely determine all the model parameters, if the Dirac
CP-violating phase $\delta$ is measured in the future neutrino
oscillation experiments.

\section{Remarks and Conclusions}

How to understand lepton mass spectra and neutrino mixing pattern
remains an open question in elementary particle physics. Flavor
symmetry is currently a powerful tool to tackle this longstanding
problem. In this paper, we apply the $S_3$ symmetry to both
charged-lepton and neutrino mass matrices. In order to explain
realistic lepton mass spectra and neutrino mixing angles, the $S_3$
symmetry is explicitly broken down via $S_3 \to Z_3 \to \emptyset$
in the charged-lepton sector, while via $S_3 \to Z_2 \to \emptyset$
in the neutrino sector. Along this line, the mass matrices of
charged leptons and neutrinos are constructed step by step. Some
interesting features of this model have been explored:
\begin{itemize}
\item It seems reasonable that the flavor symmetry first breaks
down to its subgroups. The permutation group $S_3$ contains only two
kinds of non-trivial subgroups, i.e., $Z_3$ and $Z_2$. For the
breaking chain $S_3 \to Z_3 \to \emptyset$, the mass matrix has to
be non-symmetric, so it is only allowed for charged leptons.
Neutrinos are assumed to be Majorana particles, which are actually
realized in various seesaw models. Therefore, neutrino mass matrix
should be symmetric, which is not spoiled in the $S_3 \to Z_2 \to
\emptyset$ breaking chain.

\item After the flavor symmetry breaking, lepton mass matrices are
determined by ten parameters, i.e., $c_f$, $r_f$, $\delta_f$,
$\epsilon_f$ and $\varepsilon_f$ for $f = \ell, \nu$. It has been
shown that all of them are completely fixed by the ten observables
in the lepton sector, namely three charged-lepton masses $(m_e,
m_\mu, m_\tau)$, three neutrino masses $(m_1, m_2, m_3)$, three
neutrino mixing angles $(\theta_{12}, \theta_{23}, \theta_{13})$ and
one Dirac CP-violating phase $\delta$. If leptonic CP violation is
finally measured in the future long-baseline neutrino experiments,
the model parameters will be fully determined. In light of the
recent T2K indication of relatively-large $\theta_{13}$, the
discovery of CP violation in neutrino oscillations seems very
promising.

\item In our symmetry-breaking scheme, it has been found
that $\theta_{13}$ and $\theta_{23}$ can receive large corrections
from the $S_3$ symmetry-breaking terms. More explicitly, both
corrections from charged leptons and neutrinos are significant for
obtaining a relatively-large $\theta_{13}$, while $\theta_{23}$ is
mainly sensitive to the breaking term in the neutrino mass matrix.
We show that both a relatively-large $\theta_{13}$ and a
nearly-maximal $\theta_{23}$ can be accommodated simultaneously,
which is not the case for previous $S_3$ symmetry models.
\end{itemize}

Finally, it is worth mentioning that the $S_3 \to Z_2 \to \emptyset$
chain can be applied to charged leptons. In addition, the two-stage
breaking scheme could also be applicable to quarks, and may be
helpful in understanding quark mass spectra, mixing angles and CP
violation. It should be very interesting if the mass spectra and
mixing patterns for both quarks and leptons can be understood in
this way, and a renormalizable field-theory model with the $S_3$
symmetry can be constructed to realize the derived fermion mass
matrices. We leave these issues for future works.

\acknowledgements

The author is indebted to Zhi-zhong Xing for reading the manuscript
and helpful suggestions, and to Georg Raffelt for warm hospitality
at the Max-Planck-Institut f\"ur Physik, M\"{u}nchen. This work was
supported by the Alexander von Humboldt Foundation.

%%%%%%%%%%%%%%%%%%%%%%%%%%%%%%%%%%%%%%%%%%%%%%%%%%%%%%%%%%%%%%%%%%%%%%
%%%  Bibliography  %%%%%%%%%%%%%%%%%%%%%%%%%%%%%%%%%%%%%%%%%%%%%%%%%%%
%%%%%%%%%%%%%%%%%%%%%%%%%%%%%%%%%%%%%%%%%%%%%%%%%%%%%%%%%%%%%%%%%%%%%%

%%%%%%%%%%%%%%%%%%%%%%%%%%%%%%%%%%%%%%%%%%%%%%%%%%%%%%%%%%%%%%%%%%%%%%

\begin{thebibliography}{00}
%\cite{Ishimori:2010au}
\bibitem{Ishimori:2010au}
  H.~Ishimori, T.~Kobayashi, H.~Ohki, Y.~Shimizu, H.~Okada and M.~Tanimoto,
  %``Non-Abelian Discrete Symmetries in Particle Physics,''
  Prog.\ Theor.\ Phys.\ Suppl.\  {\bf 183}, 1 (2010)
  [arXiv:1003.3552 [hep-th]].
  %%CITATION = PTPSA,183,1;%%

%\cite{Ma:2001dn}
\bibitem{Ma:2001dn}
  E.~Ma and G.~Rajasekaran,
  %``Softly broken A(4) symmetry for nearly degenerate neutrino masses,''
  Phys.\ Rev.\  D {\bf 64}, 113012 (2001)
  [arXiv:hep-ph/0106291].
  %%CITATION = PHRVA,D64,113012;%%

%\cite{Ma:2002ge}
\bibitem{Ma:2002ge}
  E.~Ma,
  %``Plato's fire and the neutrino mass matrix,''
  Mod.\ Phys.\ Lett.\  A {\bf 17}, 2361 (2002)
  [arXiv:hep-ph/0211393].
  %%CITATION = MPLAE,A17,2361;%%

%\cite{Altarelli:2005yp}
\bibitem{Altarelli:2005yp}
  G.~Altarelli and F.~Feruglio,
  %``Tri-bimaximal neutrino mixing from discrete symmetry in extra dimensions,''
  Nucl.\ Phys.\  B {\bf 720}, 64 (2005)
  [arXiv:hep-ph/0504165].
  %%CITATION = NUPHA,B720,64;%%

%\cite{Altarelli:2005yx}
\bibitem{Altarelli:2005yx}
  G.~Altarelli and F.~Feruglio,
  %``Tri-bimaximal neutrino mixing, A(4) and the modular symmetry,''
  Nucl.\ Phys.\  B {\bf 741}, 215 (2006)
  [arXiv:hep-ph/0512103].
  %%CITATION = NUPHA,B741,215;%%

%\cite{Yamanaka:1981pa}
\bibitem{Yamanaka:1981pa}
  Y.~Yamanaka, H.~Sugawara and S.~Pakvasa,
  %``PERMUTATION SYMMETRIES AND THE FERMION MASS MATRIX,''
  Phys.\ Rev.\  D {\bf 25}, 1895 (1982)
  [Erratum-ibid.\  D {\bf 29}, 2135 (1984)].
  %%CITATION = PHRVA,D25,1895;%%

%\cite{Brown:1984dk}
\bibitem{Brown:1984dk}
  T.~Brown, S.~Pakvasa, H.~Sugawara and Y.~Yamanaka,
  %``NEUTRINO MASSES, MIXING AND OSCILLATIONS IN S(4) MODEL OF PERMUTATION
  %SYMMETRY,''
  Phys.\ Rev.\  D {\bf 30}, 255 (1984).
  %%CITATION = PHRVA,D30,255;%%

%\cite{Brown:1984mq}
\bibitem{Brown:1984mq}
  T.~Brown, N.~Deshpande, S.~Pakvasa and H.~Sugawara,
  %``CP NONCONSERVATION AND RARE PROCESSES IN S(4) MODEL OF PERMUTATION
  %SYMMETRY,''
  Phys.\ Lett.\  B {\bf 141}, 95 (1984).
  %%CITATION = PHLTA,B141,95;%%

%\cite{Ma:2005pd}
\bibitem{Ma:2005pd}
  E.~Ma,
  %``Neutrino mass matrix from S(4) symmetry,''
  Phys.\ Lett.\  B {\bf 632}, 352 (2006)
  [arXiv:hep-ph/0508231].
  %%CITATION = PHLTA,B632,352;%%

%\cite{Hagedorn:2006ug}
\bibitem{Hagedorn:2006ug}
  C.~Hagedorn, M.~Lindner and R.~N.~Mohapatra,
  %``S(4) flavor symmetry and fermion masses: Towards a grand unified theory of
  %flavor,''
  JHEP {\bf 0606}, 042 (2006)
  [arXiv:hep-ph/0602244].
  %%CITATION = JHEPA,0606,042;%%

%\cite{Zhang:2006fv}
\bibitem{Zhang:2006fv}
  H.~Zhang,
  %``Flavor S(4) x Z(2) symmetry and neutrino mixing,''
  Phys.\ Lett.\  B {\bf 655}, 132 (2007)
  [arXiv:hep-ph/0612214].
  %%CITATION = PHLTA,B655,132;%%

%\cite{Lam:2008sh}
\bibitem{Lam:2008sh}
  C.~S.~Lam,
  %``The Unique Horizontal Symmetry of Leptons,''
  Phys.\ Rev.\  D {\bf 78}, 073015 (2008)
  [arXiv:0809.1185 [hep-ph]].
  %%CITATION = PHRVA,D78,073015;%%

%\cite{Ding:2009iy}
\bibitem{Ding:2009iy}
  G.~J.~Ding,
  %``Fermion Masses and Flavor Mixings in a Model with S(4) Flavor Symmetry,''
  Nucl.\ Phys.\  B {\bf 827}, 82 (2010)
  [arXiv:0909.2210 [hep-ph]].
  %%CITATION = NUPHA,B827,82;%%


%\cite{Yang:2011fh}
\bibitem{Yang:2011fh}
  R.~Z.~Yang and H.~Zhang,
  %``Minimal seesaw model with S_4 flavor symmetry,''
  Phys.\ Lett.\  B {\bf 700}, 316 (2011)
  [arXiv:1104.0380 [hep-ph]].
  %%CITATION = PHLTA,B700,316;%%

%\cite{Harrison:2002er}
\bibitem{Harrison:2002er}
  P.~F.~Harrison, D.~H.~Perkins and W.~G.~Scott,
  %``Tri-bimaximal mixing and the neutrino oscillation data,''
  Phys.\ Lett.\  B {\bf 530}, 167 (2002)
  [arXiv:hep-ph/0202074].
  %%CITATION = PHLTA,B530,167;%%

%\cite{Xing:2002sw}
\bibitem{Xing:2002sw}
  Z.~Z.~Xing,
  %``Nearly tri bimaximal neutrino mixing and CP violation,''
  Phys.\ Lett.\  B {\bf 533}, 85 (2002)
  [arXiv:hep-ph/0204049].
  %%CITATION = PHLTA,B533,85;%%

%\cite{Harrison:2002kp}
\bibitem{Harrison:2002kp}
  P.~F.~Harrison and W.~G.~Scott,
  %``Symmetries and generalizations of tri - bimaximal neutrino mixing,''
  Phys.\ Lett.\  B {\bf 535}, 163 (2002)
  [arXiv:hep-ph/0203209].
  %%CITATION = PHLTA,B535,163;%%

%\cite{He:2003rm}
\bibitem{He:2003rm}
  X.~G.~He and A.~Zee,
  %``Some simple mixing and mass matrices for neutrinos,''
  Phys.\ Lett.\  B {\bf 560}, 87 (2003)
  [arXiv:hep-ph/0301092].
  %%CITATION = PHLTA,B560,87;%%


%\cite{Abe:2011sj}
\bibitem{Abe:2011sj}
   K. Abe, {\it et al.}  [T2K Collaboration],
  %``Indication of Electron Neutrino Appearance from an Accelerator-produced
  %Off-axis Muon Neutrino Beam,''
  Phys.\ Rev.\ Lett.\  {\bf 107}, 041801 (2011)
  [arXiv:1106.2822 [hep-ex]].
  %%CITATION = ARXIV:1106.2822;%%

%\cite{Fogli:2008jx}
\bibitem{Fogli:2008jx}
  G.~L.~Fogli, E.~Lisi, A.~Marrone, A.~Palazzo and A.~M.~Rotunno,
  %``Hints of theta(13) > 0 from global neutrino data analysis,''
  Phys.\ Rev.\ Lett.\  {\bf 101}, 141801 (2008)
  [arXiv:0806.2649 [hep-ph]].
  %%CITATION = PRLTA,101,141801;%%

%\cite{GonzalezGarcia:2010er}
\bibitem{GonzalezGarcia:2010er}
  M.~C.~Gonzalez-Garcia, M.~Maltoni and J.~Salvado,
  %``Updated global fit to three neutrino mixing: status of the hints of theta13
  %> 0,''
  JHEP {\bf 1004}, 056 (2010)
  [arXiv:1001.4524 [hep-ph]].
  %%CITATION = JHEPA,1004,056;%%

%\cite{Schwetz:2011qt}
\bibitem{Schwetz:2011qt}
  T.~Schwetz, M.~Tortola and J.~W.~F.~Valle,
  %``Global neutrino data and recent reactor fluxes: status of three-flavour
  %oscillation parameters,''
  New J.\ Phys.\  {\bf 13}, 063004 (2011)
  [arXiv:1103.0734 [hep-ph]].
  %%CITATION = NJOPF,13,063004;%%

%\cite{Shimizu:2011xg}
\bibitem{Shimizu:2011xg}
  Y.~Shimizu, M.~Tanimoto and A.~Watanabe,
  %``Breaking Tri-bimaximal Mixing and Large $\theta_{13}$,''
  Prog.\ Theor.\ Phys.\ {\bf 126}, 81 (2011)
  [arXiv:1105.2929 [hep-ph]].
  %%CITATION = ARXIV:1105.2929;%%

%\cite{King:2011xu}
\bibitem{King:2011xu}
  S.~F.~King,
  %``Neutrino Mass Models: Impact of non-zero reactor angle,''
  arXiv:1106.4239 [hep-ph].
  %%CITATION = ARXIV:1106.4239;%%

%\cite{Ma:2011yi}
\bibitem{Ma:2011yi}
  E.~Ma and D.~Wegman,
  %``Nonzero theta(13) for neutrino mixing in the context of A(4) symmetry,''
  Phys.\ Rev.\ Lett.\  {\bf 107}, 061803 (2011)
  [arXiv:1106.4269 [hep-ph]].
  %%CITATION = ARXIV:1106.4269;%%

%\cite{He:2011gb}
\bibitem{He:2011gb}
  X.~G.~He and A.~Zee,
  %``Minimal Modification to Tri-bimaximal Mixing,''
  arXiv:1106.4359 [hep-ph].
  %%CITATION = ARXIV:1106.4359;%%

%\cite{Fritzsch:1995dj}
\bibitem{Fritzsch:1995dj}
  H.~Fritzsch and Z.~Z.~Xing,
  %``Lepton mass hierarchy and neutrino oscillations,''
  Phys.\ Lett.\  B {\bf 372}, 265 (1996)
  [arXiv:hep-ph/9509389].
  %%CITATION = PHLTA,B372,265;%%

%\cite{Fritzsch:1998xs}
\bibitem{Fritzsch:1998xs}
  H.~Fritzsch and Z.~Z.~Xing,
  %``Large leptonic flavor mixing and the mass spectrum of leptons,''
  Phys.\ Lett.\  B {\bf 440}, 313 (1998)
  [arXiv:hep-ph/9808272].
  %%CITATION = PHLTA,B440,313;%%

%\cite{Fukugita:1998vn}
\bibitem{Fukugita:1998vn}
  M.~Fukugita, M.~Tanimoto and T.~Yanagida,
  %``Atmospheric neutrino oscillation and a phenomenological lepton mass
  %matrix,''
  Phys.\ Rev.\  D {\bf 57}, 4429 (1998)
  [arXiv:hep-ph/9709388].
  %%CITATION = PHRVA,D57,4429;%%

%\cite{Koide:1999mx}
\bibitem{Koide:1999mx}
  Y.~Koide,
  %``Universal seesaw mass matrix model with an S(3) symmetry,''
  Phys.\ Rev.\  D {\bf 60}, 077301 (1999)
  [arXiv:hep-ph/9905416].
  %%CITATION = PHRVA,D60,077301;%%

%\cite{Fritzsch:1999im}
\bibitem{Fritzsch:1999im}
  H.~Fritzsch and Z.~Z.~Xing,
  %``Maximal neutrino mixing and maximal CP violation,''
  Phys.\ Rev.\  D {\bf 61}, 073016 (2000)
  [arXiv:hep-ph/9909304].
  %%CITATION = PHRVA,D61,073016;%%

%\cite{Tanimoto:2000fz}
\bibitem{Tanimoto:2000fz}
  M.~Tanimoto,
  %``Large mixing angle MSW solution in S(3) flavor symmetry,''
  Phys.\ Lett.\  B {\bf 483}, 417 (2000)
  [arXiv:hep-ph/0001306].
  %%CITATION = PHLTA,B483,417;%%

%\cite{Branco:2001hn}
\bibitem{Branco:2001hn}
  G.~C.~Branco and J.~I.~Silva-Marcos,
  %``The Symmetry behind extended flavor democracy and large leptonic mixing,''
  Phys.\ Lett.\  B {\bf 526}, 104 (2002)
  [arXiv:hep-ph/0106125].
  %%CITATION = PHLTA,B526,104;%%

%\cite{Fujii:2002jw}
\bibitem{Fujii:2002jw}
  M.~Fujii, K.~Hamaguchi and T.~Yanagida,
  %``Leptogenesis with almost degenerate majorana neutrinos,''
  Phys.\ Rev.\  D {\bf 65}, 115012 (2002)
  [arXiv:hep-ph/0202210].
  %%CITATION = PHRVA,D65,115012;%%

%\cite{Kubo:2003iw}
\bibitem{Kubo:2003iw}
  J.~Kubo, A.~Mondragon, M.~Mondragon and E.~Rodriguez-Jauregui,
  %``The Flavor symmetry,''
  Prog.\ Theor.\ Phys.\  {\bf 109}, 795 (2003)
  [Erratum-ibid.\  {\bf 114}, 287 (2005)]
  [arXiv:hep-ph/0302196].
  %%CITATION = PTPKA,109,795;%%

%\cite{Chen:2004rr}
\bibitem{Chen:2004rr}
  S.~L.~Chen, M.~Frigerio and E.~Ma,
  %``Large neutrino mixing and normal mass hierarchy: A Discrete
  %understanding,''
  Phys.\ Rev.\  D {\bf 70}, 073008 (2004)
  [Erratum-ibid.\  D {\bf 70}, 079905 (2004)]
  [arXiv:hep-ph/0404084].
  %%CITATION = PHRVA,D70,073008;%%

%\cite{Fritzsch:2004xc}
\bibitem{Fritzsch:2004xc}
  H.~Fritzsch and Z.~Z.~Xing,
  %``Democratic neutrino mixing reexamined,''
  Phys.\ Lett.\  B {\bf 598}, 237 (2004)
  [arXiv:hep-ph/0406206].
  %%CITATION = PHLTA,B598,237;%%

%\cite{Grimus:2005mu}
\bibitem{Grimus:2005mu}
  W.~Grimus and L.~Lavoura,
  %``S(3) x Z(2) model for neutrino mass matrices,''
  JHEP {\bf 0508}, 013 (2005)
  [arXiv:hep-ph/0504153].
  %%CITATION = JHEPA,0508,013;%%

%\cite{Mohapatra:2006pu}
\bibitem{Mohapatra:2006pu}
  R.~N.~Mohapatra, S.~Nasri and H.~B.~Yu,
  %``S(3) symmetry and tri-bimaximal mixing,''
  Phys.\ Lett.\  B {\bf 639}, 318 (2006)
  [arXiv:hep-ph/0605020].
  %%CITATION = PHLTA,B639,318;%%

%\cite{Jora:2006dh}
\bibitem{Jora:2006dh}
  R.~Jora, S.~Nasri and J.~Schechter,
  %``An Approach to permutation symmetry for the electroweak theory,''
  Int.\ J.\ Mod.\ Phys.\  A {\bf 21}, 5875 (2006)
  [arXiv:hep-ph/0605069].
  %%CITATION = IMPAE,A21,5875;%%

%\cite{Chen:2007zj}
\bibitem{Chen:2007zj}
  C.~Y.~Chen and L.~Wolfenstein,
  %``Consequences of approximate S(3) symmetry of the neutrino mass matrix,''
  Phys.\ Rev.\  D {\bf 77}, 093009 (2008)
  [arXiv:0709.3767 [hep-ph]].
  %%CITATION = PHRVA,D77,093009;%%

%\cite{Jora:2009gz}
\bibitem{Jora:2009gz}
  R.~Jora, J.~Schechter and M.~Naeem Shahid,
  %``Perturbed S(3) neutrinos,''
  Phys.\ Rev.\  D {\bf 80}, 093007 (2009)
  [Erratum-ibid.\  D {\bf 82}, 079902 (2010)]
  [arXiv:0909.4414 [hep-ph]].
  %%CITATION = PHRVA,D80,093007;%%

%\cite{Dicus:2010iq}
\bibitem{Dicus:2010iq}
  D.~A.~Dicus, S.~F.~Ge and W.~W.~Repko,
  %``Neutrino mixing with broken $S_3$ symmetry,''
  Phys.\ Rev.\  D {\bf 82}, 033005 (2010)
  [arXiv:1004.3266 [hep-ph]].
  %%CITATION = PHRVA,D82,033005;%%

%\cite{Xing:2010iu}
\bibitem{Xing:2010iu}
  Z.~Z.~Xing, D.~Yang and S.~Zhou,
  %``Broken S_3 Flavor Symmetry of Leptons and Quarks: Mass Spectra and Flavor
  %Mixing Patterns,''
  Phys.\ Lett.\  B {\bf 690}, 304 (2010)
  [arXiv:1004.4234 [hep-ph]].
  %%CITATION = PHLTA,B690,304;%%

%\cite{Jora:2010at}
\bibitem{Jora:2010at}
  R.~Jora, J.~Schechter and M.~N.~Shahid,
  %``Doubly perturbed $S_{3}$ neutrinos and the $s_{13}$ mixing parameter,''
  Phys.\ Rev.\  D {\bf 82}, 053006 (2010)
  [arXiv:1006.3307 [hep-ph]].
  %%CITATION = PHRVA,D82,053006;%%

%\cite{Dev:2011qy}
\bibitem{Dev:2011qy}
  S.~Dev, S.~Gupta and R.~R.~Gautam,
  %``Broken $S_3$ Symmetry in the Neutrino Mass Matrix,''
  Phys.\ Lett.\  B {\bf 702}, 28 (2011)
  [arXiv:1106.3873 [hep-ph]].
  %%CITATION = PHLTA,B702,28;%%

%\cite{Xing:2010pn}
\bibitem{Xing:2010pn}
  Z.~Z.~Xing,
  %``A Shift from Democratic to Tri-bimaximal Neutrino Mixing with Relatively
  %Large theta_{13},''
  Phys.\ Lett.\  B {\bf 696}, 232 (2011)
  [arXiv:1011.2954 [hep-ph]].
  %%CITATION = PHLTA,B696,232;%%

%\cite{Xing:2011at}
\bibitem{Xing:2011at}
  Z.~Z.~Xing,
  %``The T2K Indication of Relatively Large theta_13 and a Natural Perturbation
  %to the Democratic Neutrino Mixing Pattern,''
  arXiv:1106.3244 [hep-ph].
  %%CITATION = ARXIV:1106.3244;%%

%\cite{Xing:2011zza}
\bibitem{Xing:2011zza}
  Z.~Z.~Xing and S.~Zhou,
  ``Neutrinos in particle physics, astronomy and cosmology," {\it Zhejiang
  University Press and Springer-Verlag (2011) 426 p}.
  %%CITATION = ISBN-978-3-642-17559-6;%%

%\cite{Fukuyama:1997ky}
\bibitem{Fukuyama:1997ky}
  T.~Fukuyama and H.~Nishiura,
  %``Mass matrix of Majorana neutrinos,''
  arXiv:hep-ph/9702253.
  %%CITATION = HEP-PH/9702253;%%

%\cite{Xing:2006xa}
\bibitem{Xing:2006xa}
  Z.~Z.~Xing, H.~Zhang and S.~Zhou,
  %``Nearly Tri-bimaximal Neutrino Mixing and CP Violation from mu-tau Symmetry
  %Breaking,''
  Phys.\ Lett.\  B {\bf 641}, 189 (2006)
  [arXiv:hep-ph/0607091]; and references therein.
  %%CITATION = PHLTA,B641,189;%%

%\cite{pdg}
\bibitem{pdg} K. Nakamura, {\it et al.} (Particle Data Group),
J. Phys. G {\bf 37}, 075021 (2010).

%\cite{Xing:2007fb}
\bibitem{Xing:2007fb}
  Z.~Z.~Xing, H.~Zhang and S.~Zhou,
  %``Updated Values of Running Quark and Lepton Masses,''
  Phys.\ Rev.\  D {\bf 77}, 113016 (2008)
  [arXiv:0712.1419 [hep-ph]].
  %%CITATION = PHRVA,D77,113016;%%

\end{thebibliography}
\end{document}